\documentstyle[12pt]{article}

\def\simlt{\stackrel{<}{{}_\sim}}
\def\simgt{\stackrel{>}{{}_\sim}}
\def\be{\begin{equation}}
\def\ee{\end{equation}}
\def\bea{\begin{eqnarray}}
\def\eea{\end{eqnarray}}
\def\simlt{\stackrel{<}{{}_\sim}}
\def\simgt{\stackrel{>}{{}_\sim}}
\def\ap#1#2#3   {{\em Ann. Phys. (NY)} {\bf#1} (#2) #3.}
\def\apj#1#2#3  {{\em  Astrophys. J.} {\bf#1} (#2) #3.}
\def\apjl#1#2#3 {{\em Astrophys. J. Lett.} {\bf#1} (#2) #3.}
\def\app#1#2#3  {{\em Acta. Phys. Pol.} {\bf#1} (#2) #3.}
\def\ar#1#2#3   {{\em Ann. Rev. Nucl. Part. Sci.} {\bf#1} (#2) #3.}
\def\cpc#1#2#3  {{\em Computer Phys. Comm.} {\bf#1} (#2) #3.}
\def\err#1#2#3  {{\em Erratum} {\bf#1} (#2) #3.}
\def\ib#1#2#3   {{\em ibid.} {\bf#1} (#2) #3.}
\def\jmp#1#2#3  {{\em J. Math. Phys.} {\bf#1} (#2) #3.}
\def\ijmp#1#2#3 {{\em Int. J. Mod. Phys.} {\bf#1} (#2) #3.}
\def\jetp#1#2#3 {{\em JETP Lett.} {\bf#1} (#2) #3.}
\def\jpg#1#2#3  {{\em J. Phys. G.} {\bf#1} (#2) #3.}
\def\mpl#1#2#3  {{\em Mod. Phys. Lett.} {\bf#1} (#2) #3.}
\def\nat#1#2#3  {{\em Nature (London)} {\bf#1} (#2) #3.}
\def\nc#1#2#3   {{\em Nuovo Cim.} {\bf#1} (#2) #3.}
\def\nim#1#2#3  {{\em Nucl. Instr. Meth.} {\bf#1} (#2) #3.}
\def\np#1#2#3   {{\em Nucl. Phys.} {\bf#1} (#2) #3.}
\def\pcps#1#2#3 {{\em Proc. Cam. Phil. Soc.} {\bf#1} (#2) #3.}
\def\pl#1#2#3   {{\em Phys. Lett.} {\bf#1} (#2) #3.}
\def\prep#1#2#3 {{\em Phys. Rep.} {\bf#1} (#2) #3.}
\def\prev#1#2#3 {{\em Phys. Rev.} {\bf#1} (#2) #3.}
\def\prl#1#2#3  {{\em Phys. Rev. Lett.} {\bf#1} (#2) #3.}
\def\prs#1#2#3  {{\em Proc. Roy. Soc.} {\bf#1} (#2) #3.}
\def\ptp#1#2#3  {{\em Prog. Th. Phys.} {\bf#1} (#2) #3.}
\def\ps#1#2#3   {{\em Physica Scripta} {\bf#1} (#2) #3.}
\def\rmp#1#2#3  {{\em Rev. Mod. Phys.} {\bf#1} (#2) #3.}
\def\rpp#1#2#3  {{\em Rep. Prog. Phys.} {\bf#1} (#2) #3.}
\def\sjnp#1#2#3 {{\em Sov. J. Nucl. Phys.} {\bf#1} (#2) #3.}
\def\spj#1#2#3  {{\em Sov. Phys. JEPT} {\bf#1} (#2) #3.}
\def\spu#1#2#3  {{\em Sov. Phys.-Usp.} {\bf#1} (#2) #3.}
\def\zp#1#2#3   {{\em Zeit. Phys.} {\bf#1} (#2) #3.}
\def\NPB#1#2#3{{\it Nucl.~Phys.} {\bf{B#1}} (19#2) #3}
\def\PLB#1#2#3{{\it Phys.~Lett.} {\bf{B#1}} (19#2) #3}
\def\PRD#1#2#3{{\it Phys.~Rev.} {\bf{D#1}} (19#2) #3}
\def\PRL#1#2#3{{\it Phys.~Rev.~Lett.} {\bf{#1}} (19#2) #3}

\def\beq{\begin{equation}}
\def\eeq{\end{equation}}
\def\bea{\begin{eqnarray}}
\def\eea{\end{eqnarray}}

\begin{document}
\begin{flushright}
\vspace{-0.3cm}
{\normalsize CERN-TH/99-261}\\
{\normalsize ANL-HEP-PR-99-86}\\ 
\end{flushright}
\begin{center}
{\Large \bf
Cosmological Magnetic Fields From Gauge-Mediated \\
\vskip 0.5cm 
Supersymmetry-Breaking Models} \\
\vskip 1.5cm
{Alejandra Kandus$^1$, $\;$ Esteban A. Calzetta$^1$, \\
~\\ 
Francisco D. Mazzitelli$^1$  and   Carlos E.M. Wagner$^{2,3}$}\\
~\\
~\\
{{\it $^1$Departamento de F\'{\i}sica, Fac. Cs. Ex. y Nat. UBA,}}\\
{{\it Ciudad Universitaria, (1428) Buenos Aires, Argentina}}\\
~\\
{{\it $^2$Theory Division, CERN, CH-1211 Geneva 23, Switzerland}} \\
~\\
{{\it $^3$ High Energy Division, Argonne National Laboratory,
Argonne, IL 60439, USA}} \\
\end{center}
\begin{quote}
\begin{center}
{\bf Abstract}
\end{center}
We study the generation of primordial magnetic fields, coherent over
cosmologically interesting scales, by gravitational creation of charged 
scalar particles during the reheating period. We show that magnetic fields
consistent with those detected by observation may be obtained if
the particle mean life $\tau_s$ is in the range $10^{-14}$ sec $
\simlt \tau_s \simlt 
10^{-7}$ sec. We apply this mechanism to minimal gauge-mediated 
supersymmetry-breaking models, in the case in which the 
lightest stau $\tilde{\tau}_1$ is the next-to-lightest supersymmetric
particle. We show that, for a large range of phenomenologically
acceptable values of the supersymmetry-breaking scale $\sqrt{F}$, the
generated primordial magnetic field can be strong enough to
seed the galactic dynamo. 
\end{quote}
\vskip 1.5cm 
\begin{flushleft} 
{\normalsize CERN-TH/99-261}\\
%
\end{flushleft} 
\newpage
Quite homogeneous magnetic
fields of intensity 
$B\simeq 3\times 10^{-6}$ Gauss are present in all structures 
of our Universe: galaxies, galaxy clusters
and hydrogen clouds \cite{zeldovich,kronberg,wolfe,oren}. 
One of the mechanisms for the generation of
magnetic fields is the primordial
generation of a seed field that is further amplified by gravitational collapse 
and/or dynamo \cite{zeldovich}. For the origin of the seed field, 
several mechanisms have been proposed recently: it 
has been suggested that a primordial field might be produced during
the inflationary period if conformal invariance 
is broken \cite{widrow,dolgov}; in superstring-inspired models, 
the coupling between the electromagnetic field 
and the dilaton breaks
conformal invariance and might produce the seed field 
\cite{ratra,veneziano}; gauge-invariant
couplings between the electromagnetic 
field and the space-time curvature also break
conformal invariance, but produces in general an uninterestingly 
small seed field~\cite{spedalieri}; other mechanisms are based on,
for example, 
first order cosmological phase transitions
and on the existence of 
topological defects \cite{olinto,otros,shaposhnikov}.

Recently,
a new mechanism for cosmological magnetic field generation
 was proposed \cite{ckm}, based on the presence
during inflation of a charged, minimally coupled scalar field in
its invariant vacuum state \cite{allen}. When the transition to radiation
takes place, quantum creation of charged particles occurs because of the
release of gravitational energy. The mean electric current is zero, 
but stochastic fluctuations
around that mean give a non-vanishing contribution. The magnetic field 
induced by this stochastic current was 
sufficient to seed the galactic dynamo. However, there remained the 
important issue of finding a suitable scalar particle to 
generate the electric current source of the magnetic field. 

   In this letter we address this problem in the context of 
gauge-mediated supersymmetry-breaking models (GMSB). 
In the simplest version of these
models, supersymmetry-breaking is communicated to the visible
sector through a set of massive fields, called messengers, which
carry non-trivial quantum numbers under the gauge group~\cite{gms,at1}. 
The messengers $\Phi_I$, $\bar{\Phi}_I$ are assumed to 
acquire an explicit mass $M_I$ by the vacuum expectation value 
of a singlet field 
$\langle X_I \rangle = M_I$, via a superpotential coupling
\bea
W = X_I \bar{\Phi}_I \Phi_I.
\eea
A vacuum expectation value of the auxiliary component $F_I$ of the
field $X_I$ breaks supersymmetry and induces, through the gauge interactions
of the messenger fields, the supersymmetry-breaking masses in
the observable sector. For the simplest case of $N$ sets of
messenger fields belonging to the fundamental representation of
SU(5) and a single field $X$, one gets gaugino masses 
\bea
M_i  = \frac{N \alpha_i}{4\pi} \frac{F}{M}.
\eea
where $i = 1,2$,3 are associated with the gauge groups $U(1)_Y,SU(2)$ and 
$SU(3)_c$, respectively.

The scalar masses not affected by Yukawa couplings are
given by
\bea
m_S^2(\mu) = \frac{2 \;N \;c_S^i}{16 \pi^2} \alpha_i^2(0) 
\left(\frac{F}{M}\right)^2
 - \frac{2 c_S^i}{b_i} M_i^2(0) 
\left( \frac{ \alpha_i^2(\mu) - \alpha_i^2(0)}
{\alpha_i^2(0)} \right),
\eea 
where  $m_S$ are the supersymmetry-breaking masses for
gauginos and scalars, respectively, $\mu$ is the renormalization
group scale with $\mu =0$ being identified with the messenger mass
scale, $c_S^i$ is the quadratic 
Casimir of the scalar particle under the $i$-gauge group, and
$\alpha_i$ and $b_i$ are
the corresponding gauge coupling and MSSM beta 
function coefficients~\footnote{For more general expressions see,
for instance, Ref.~\cite{wagner}}. From the above, it is easy to 
see that the right-handed sleptons
are the lightest scalars in the spectrum and, for
$N > 1$, the
lighest stau can easily become lighter than the lightest neutralino.
The lightest stau can also become lighter than the lightest
neutralino due to mixing effects, for moderate and large 
values of $\tan\beta$, for any value of $N$.
For the characteristic values of the
supersymmetry-breaking scale $F$, however, the lightest supersymmetric
particle is the gravitino. Indeed, the gravitino mass is given by
\bea
m_{\tilde G} = \frac{F}{ \sqrt{3} M_{Pl}},
\eea
where $M_{Pl}$ is the Planck scale
(we are identifying $F$ with the fundamental supersymmetry-breaking 
scale $F_0$). Hence, the gravitino is the
lightest supersymmetric particle for any messenger mass
$M$ much lower than the GUT scale.

In general, under the assumption
of R-parity conservation~\cite{rpgm}, 
the next-to-lightest SUSY particle  will decay
into a gravitino and a standard particle with an inverse 
decay rate~\cite{decay}
\begin{equation}
\tau = \frac 1{\tilde k^2} \left( \frac{100\; {\rm GeV}}
{m_{\rm NLSP}}\right) ^5
\left( \frac{\sqrt{F}}{100 
{\rm TeV}} \right) ^4 3\times 10^{11} {\rm GeV}^{-1}, 
\label{ev}
\end{equation}
where $m_{\rm NLSP}$ is the mass of the NLSP particle and 
$\tilde k$ is a projection factor equal to the component in the NLSP of
the superpartner of the particle the NLSP is decaying into.
For the
case of the stau decaying into a tau and a gravitino, $\tilde k = 1$.

Constraints on the value of the supersymmetry-breaking scale
may be obtained, for example, by the requirement that the gravitino density
does not overclose the Universe. For instance,
if the gravitino mass $m_{\tilde{G}} > 1$ keV, 
the temperature at the beginning of the radiation-dominated epoch, 
called {\sl the reheat temperature} $T_{\gamma }$, should
be much smaller than the GUT scale in order to avoid overproduction
of gravitinos~\cite{gherghetta98}. The exact bound 
on $T_{\gamma}$ depends on the gravitino mass. For relatively large
values of the gravitino mass, corresponding to $\sqrt{F} \simeq
10^{9}$~GeV, ($M \simeq 10^{13}$ GeV), an upper bound
on $T_{\gamma}$ of the order of $10^7$ GeV is 
obtained~\footnote{For large values of the gravitino mass,
the non-thermal production of gravitinos tends to be
dominant, and induces a tighter bound on the reheat temperature,
which may be of the order of the weak scale~\cite{GTR}.}. 
The bound becomes even smaller for smaller values of $F$. 
On the other hand, for $m_{\tilde{G}} < 1$~keV and for
any value of the reheat temperature larger than the weak
scale, the gravitinos will be in thermal equilibrium at early
times and, for these range of masses, the gravitinos 
are sufficiently light  to lead to cosmologically acceptable values 
of the relic density.  

The reheating period can be characterized by the temperature $T_I$ obtained
by thermalization after preheating and $T_{\gamma}$, the temperature at
the beginning of the radiation dominated epoch~\cite{khlebnikov}.
In Ref.~\cite{ckm} an inflationary model with instantaneous reheating
was considered. In the more realistic case in which 
the reheating is extended in time, 
the number of particles created at the two main transitions, namely
inflation-reheating and reheating--radiation, as well 
as during the reheating period itself should be calculated.

We will work in conformal time, which is given by $d\eta =  dt/a(t)$. Defining
$\tau = H\eta $, where $H$ is the 
Hubble constant during inflation, and assuming that
during reheating the Universe is matter-dominated \cite{kolb}, 
the 
scale factors for the different epochs of the Universe read
\begin{eqnarray}
{\rm inflation} \;\; a_I\left( \tau \right) &=& 
\frac 1{\left( 1 - \tau \right)}\label{ai} \\
{\rm reheating} \;\; a_R\left( \tau \right) &=& \left[ 1 + 
\frac {\tau }2 \right] ^2\label{ak}\\
{\rm radiation} \;\; a_{\gamma }\left( \tau \right) &=& 
\left( \frac {T_I}{T_{\gamma }}\right) ^{1/2b} 
\left[ \tau + 2 - 
\left( \frac {T_I}{T_{\gamma }} \right) ^{1/2b}\right] . \label{am}
\end{eqnarray}
$T_I$ and $T_{\gamma }$ are the temperatures of the Universe at the
beginning of reheating and at the beginning of radiation, 
respectively. We have assumed that 
during radiation the temperature of the Universe scales 
with $a\left( \tau \right) $ as
$T \propto a\left( \tau \right) ^{-1} $ while during reheating it goes as 
$T \propto a\left( \tau \right) ^{-b}$, with $0<b<1$~\cite{Toni}.

The evolution of a charged scalar 
field is given by the Klein--Gordon equation. If
we expand the real and imaginary parts of the field as 
$\left( 2\pi \right) ^{-3/2}\int d^3 \kappa
 \phi _{\kappa } \left( \tau \right) e^{i\vec \kappa .\vec r} +$ h.c. , 
the field
equation reads
\begin{equation}
\left[ \frac{\partial ^2}{\partial \tau ^2} + k^2 + 
\left(\frac{m}{H}\right)^2 a^2\left( \tau \right) - 
\left( 1 - 6\xi\right)\frac {\ddot a \left( \tau \right)}{a\left( \tau \right)}
\right] \phi_{\kappa }\left( \tau \right) = 0 , \label{bb}
\end{equation}
where $k = H^{-1}\kappa $  ($\kappa $ being the 
comoving wavenumber) and where $\xi$ is the coupling to the curvature. 
We will consider 
the mass as built up from two contributions, 
the zero-temperature  mass $m(0) \equiv m_{\tilde{\tau}}$ and the thermal 
corrections, so that we have $m^2 = m^2(0) + g
T^2\left( \tau \right) $, where $g$ is of the order of the particle gauge
coupling constants.

For the inflationary period, we do not need the thermal 
corrections, as the temperature of that period is too low 
to be important. 
However, in supergravity theories, the possible presence of a 
non-renormalizable coupling of the
inflaton field $I$ to the scalar 
fields in the K\"ahler potential~\cite{GTR}
\begin{equation}
K_{I,\phi} = - \frac{C_H}{3} \frac{1}{M_{Pl}^2} 
I^{\dagger} I \phi^{\dagger}\phi
\end{equation}
would naturally lead to a mass contribution $\delta m^2 \simeq C_H H^2$.
Hence, in general, an effective mass of the order of the Hubble constant
will be generated, although the coefficient $C_H$ may be small or
even zero in the case when the specific effective coupling is forbidden by
symmetries of the  theory~\cite{Olive}. 

The positive--frequency solution to Eq. (\ref{bb}) for inflation reads
\begin{equation}
\phi^I_k\left( \tau \right) = \frac {\sqrt{\pi }}2 \sqrt{1 - \tau }
H_{\nu }^{(1)}\left[ k\left( 1 - \tau \right) \right] \label{bg}
\end{equation}
where $H_{\nu }^{(1)}$ are the Hankel functions, with 
\bea
\nu = 
\frac 32 \sqrt{1- \frac{16}3\xi -\frac 49 
\frac{m^2}{H^2}}, \label{bbg}
\eea
where for the charecteristic values of $m(0)$ and $H$
during inflation, $m^2/H^2 \simeq C_H$.
We will assume throughout this article that the scalar field couples
minimally to the curvature,
$\xi = 0$, and that the coefficient 
$C_H \ll 1$; we will
briefly discuss the implications of different values of these quantities
at the end of this article.
For reheating and 
radiation dominance, we propose a WKB solution 
\begin{equation}
\phi _k ^{\pm} \left( \tau \right) =
\frac 1{ \sqrt{2
\vert\omega\vert\left( \tau \right) }}  
e^{\left[ \pm i\int _0 ^{\tau } \omega \left( \tau '\right) d\tau ' \right] },
\label{bh}
\end{equation}
where
\begin{eqnarray}
{\rm reheating}\;\; \omega\left( \tau \right) &=&
\sqrt{ k^2 + \frac{m(0)^2}{H^2} a_R^2\left( \tau \right) + 
g \frac{T_I^2}{H^2} 
a_R^{2(1-b)}\left( \tau \right) - 
\frac 12 a_R^{-1}\left( \tau \right) } \label{bj} \\
{\rm radiation} \;\; \omega\left( \tau \right) &=& 
\sqrt{k^2 +\frac{m(0)^2}{H^2} 
a_{\gamma }^2\left( \tau \right) +
g \frac{T_{\gamma }^2}{H^2}a_{\gamma }^2
\left( \tau _{\gamma }\right) }.  \label{bo}
\end{eqnarray}
It is important to note that 
the frequency changes from imaginary to
real values, at a certain time $\tau _c$ during reheating. 

We match the solutions to the field equation in the different epochs at the
transition between them, i.e. at the end of inflation 
and at the end of reheating. At both
times we demand continuity of the corresponding modes 
and their first time derivatives. Care
must be taken to match the WKB solutions through $\tau _c$, where 
$\omega \left( \tau _c\right) = 0$. We obtain
\begin{equation}
\phi _k \left( \tau \right) = \alpha _k \phi _{k\gamma }^+ 
\left( \tau \right) +                               
\beta _k \phi _{k\gamma }^- \left( \tau \right) ,
\label{bk}
\end{equation}
where by $ \phi _{k\gamma }$ we denote the modes during radiation and with
\begin{eqnarray}
\alpha _k &=&  -\frac{e^{i(k-\pi /4)}}{2^{3/4}} \frac 1{k^{3/2}}
\left[ i\frac{17+2\sqrt{2}}8 e^{\left( \int _0^{\tau _c} 
\vert \omega \vert d\tau '\right) }
+ \left( \frac {17 - 2\sqrt{2}}{16}\right)
e^{\left( -\int _0^{\tau _c} \vert \omega \vert d\tau '\right) }
\right] \nonumber\\
&\simeq &\frac{{\cal O}(1)}{k^{3/2}}
\label{ef}\\
\beta _k &=& \frac{e^{i(k+\pi /4)}}{2^{3/4}} \frac 1{k^{3/2}}
\left[ -i\frac{17+2\sqrt{2}}8 e^{\left( \int _0^{\tau _c} \vert 
\omega \vert d\tau '\right) }
+ \left( \frac {17 - 2\sqrt{2}}{16}\right)
e^{\left( -\int _0^{\tau _c} \vert \omega \vert d\tau '\right) }
\right] \nonumber\\
&\simeq &\frac{{\cal O}(1)}{k^{3/2} } .
\label{eg}
\end{eqnarray}
In the above, we are ignoring the effects produced by the change
of the effective Hubble constant during the inflationary period,
which results in a change of the Bogoliubov coefficients in the
far ultraviolet~\cite{Tkachev}. These effects, however, are small 
in the range of wavelengths relevant for the analysis of the 
generation of magnetic fields, $k \simlt k_{tod}$, where
$k_{tod}$ is the comoving wave number of the relevant astrophysical 
scale under study (see below).

In order to proceed with our phenomenological analysis, 
the values of $T_I,\; T_{\gamma }$  and $H$ in the 
previous expressions
must be specified. They can be related by the age of the Universe,
which can be well approximated by the duration of the 
matter-dominated epoch. This is
given by
\begin{equation}
t_{tod} \simeq \frac 2{3H} \left( \frac{T_I}{T_{\gamma }}\right) ^{3/2b}
\left( \frac{T_{\gamma }}{T_M}\right) ^2 
\left[ \left( \frac{T_M}{T_{tod}} \right) ^{3/2} -1 \right] \label{cb}
\end{equation}
where $T_{tod} \simeq 10^{-13}$ GeV is the present 
temperature of the Universe,
$T_M \simeq 1$ eV is its temperature 
at the beginning of the
matter dominated epoch and for the Hubble constant during inflation, 
we shall assume that 
$10^{11}$ GeV $\leq H \leq 10^{13}$ GeV.
{}From Eq. (\ref{cb}) we obtain
\begin{equation}
T_I \simeq T_{\gamma }^{(3-4b)/3} \left[ \frac{3H}2 T_M^{1/2}T_{tod}^{3/2}
           \times t_{tod} \right] ^{2b/3} 
           \simeq T_{\gamma}
\left[ \frac{H M_{Pl}}{T_{\gamma}^2}\right] ^{2b/3}.
 \label{cc}
\end{equation}
where the last equality stems from $t_{tod} \simeq M_{pl}/
T_{tod}^{3/2}T_M^{1/2}$~\cite{kolb}. For $H=10^{11}$~GeV we have
\bea
T_I \simeq T_{\gamma} \left(\frac{10^{15} \; {\rm GeV}}
{T_{\gamma}}
\right)^{4b/3} .
\eea
Therefore, independently  of the value of $b$ and for the values of the
cosmological parameters considered above,
the relation $T_I > T_{\gamma}$ is  fulfilled
for any value of $T_{\gamma} < 10^{15}$ GeV.

In order to compute the magnetic field, we consider the Maxwell equation
\begin{equation}
\left[ \frac{\partial ^2}{\partial \tau^2} - \nabla ^2 + 
\sigma\left( \tau \right) \frac{\partial}{\partial\tau}\right] \vec B =
\vec\nabla \times\vec j ,
\label{ccc}
\end{equation}
where $\sigma\left( \tau\right) $ is the time-dependent conductivity of
the medium and $\vec j$ is
the electric current generated by the charged scalar
particles. Although $\left< \vec j\right> = 0$, the two-point correlation
function is different from zero and produces a non-vanishing
magnetic field. This field can be expressed in terms of the two-point
function of the scalar field as (see Ref.~\cite{ckm} for details)
\begin{eqnarray}
\langle B_{\lambda }^2\rangle &=& e^2H^4\int d\tau d\tau '\int 
\frac{d\vec k d\vec k'}{\left( 2\pi\right) ^3} 
{\cal W}^2_{kk'}\left( \lambda \right)
\vert \vec k \times \vec k'\vert ^2 
G^{ret}_{\vert k+k'\vert } \left( \tau _o, \tau \right)
G^{ret}_{\vert k+k'\vert } \left( \tau _o, \tau ' \right) \nonumber \\
& \times &\left[  4 G^0_{1k} \left( \tau ,\tau '\right) 
\delta G_{1k'} \left( \tau ,\tau '\right) +
\delta G_{1k} \left( \tau ,\tau '\right)
\delta G_{1k'} \left( \tau ,\tau '\right) \right], \label{aa}
\end{eqnarray}
where
\begin{eqnarray}
G^0_{1k} \left( \tau ,\tau '\right) &=& 
\frac{\cos \Omega _k\left( \tau ,\tau '\right)}
{\sqrt{\omega \left( \tau \right) 
\omega \left( \tau '\right) }} \label{ab} \\
\delta G_{1k} \left( \tau ,\tau '\right) &=& 2\alpha _k \beta ^*_k 
\phi _{k\gamma }^+ \left( \tau \right) \phi _{k\gamma }^+ 
\left( \tau '\right) +
2\alpha ^*_k \beta _k\phi _{k\gamma }^- \left( \tau \right) \phi _{k\gamma }^- 
\left( \tau '\right) \nonumber \\
&+& 2\vert \beta _k \vert ^2 G^0_{1k} \left( \tau ,\tau '\right), \label{ac}
\end{eqnarray}
with
$\Omega \left( \tau \right) = 
\int ^{\tau } d\tau ' \omega \left( \tau '\right) $,
$\phi _{k\gamma }^+$ ($\phi _{k\gamma }^-$) the positive (negative) 
frequency modes of the scalar field during radiation dominance 
and  with $G^{ret}_{\vert k+k'\vert } 
\left( \tau _o, \tau \right) $ the retarded propagator for the 
electromagnetic field. ${\cal W}\left( \lambda \right) _{kk'}$ is the 
window function that filters scales smaller than $\lambda $. 
Eq. (\ref{aa}) then gives the magnetic energy of a field
which is homogenous over volumes of order $\lambda^3$, the intensity of
the field therefore being estimated as 
$\sqrt{\left< B_{\lambda}^2\right>}$. From now on it will be 
understood $\lambda = k_{tod}^{-1}$.
where, as mentioned above, 
$k_{tod}$ is the comoving wavenumber of the astrophysical scale
we are interested in.

Real particle propagation can be considered as such from the
moment when the frequency becomes real,
i.e. from $\tau _c$. To evaluate Eq. (\ref{aa}) we shall
proceed in the same way as in~\cite{ckm}, and consider
only the main contribution, which originates from the  
last term between brackets, 
which is quartic in the Bogoliubov coefficients and, 
within this term, from the 
non-oscillatory contributions. We perform the $k$ integration 
with the same window 
function used in \cite{ckm}, i.e. a top-hat one. 
We propagate the magnetic field during reheating and radiation dominance
until the moment of detection with the propagator given by 
the equation $\left[ \partial ^2/\partial \tau ^2 +k^2 + 
\sigma \left( \tau \right)
\partial /\partial \tau \right] G^{ret}_{\vert k+k'\vert } 
\left( \tau _o , \tau \right)
= \delta \left( \tau _o - \tau \right)$, where $\sigma 
\left( \tau \right) $ is the
electric conductivity of the Universe. After the particles decay, 
the field propagates
conformally. We assume that during all these periods the conductivity of the
Universe is given by \cite{olinto2}
\begin{equation}
\sigma \left( \tau \right) \simeq \frac{e^{-2}T}H =
\frac{\sigma _0}{\left( \tau - \tau _*\right) ^{\alpha }}, \label{ega}
\end{equation}
where $\alpha = 2b$, $\tau _* = -2 $ and $\sigma _0 = T_I/e^2H$ 
for reheating, and 
$\alpha = 1$, $\tau _* = -2 + \left( T_I/T_{\gamma }\right) ^{1/2b}$ 
and $\sigma _0 = T_I^{1/2b}T_{\gamma }^{1-1/2b}/e^2H$
for radiation dominance.

For $\tau _o \gg \tau $, we have 
\begin{equation}
G^{ret}_{\vert k + k'\vert }\left( \tau _o, \tau \right) \simeq 
-\frac{\left( \tau - \tau _* \right) ^{\alpha}}{\sigma _0} \label{egj}.
\end{equation}

The Bogoliubov coefficients are the ones given in 
Eqs. (\ref{ef}) and (\ref{eg}).
Now we are ready to evaluate the time integrals in Eq. (\ref{aa}).  It can be 
checked that the contribution from reheating is negligible 
with respect to
the one from the radiation period. Also, the mass term dominates over the
thermal correction for a particle lifetime $t_{max} > 10 ^{-14}$ sec. 
We therefore consider the time integral 
\begin{equation}
\int _{\tau _c}^{\tau _{max}}d\tau ' 
\frac{G^{ret}\left( \tau ', \tau _o \right)}{\omega 
\left( \tau '\right) } \simeq
- \frac{2e^2H^{5/2}}{T_{\gamma }m(0)}
\left( \frac{T_{\gamma }}{T_I} \right) ^{5/4b}
\left[ t_{max} + 
\frac{1}{2H} \left( \frac{T_I}{T_{\gamma }} \right)^{3/2b}\right] ^{1/2} 
\label{ep}
\end{equation}
Now we are ready to evaluate the magnetic field. 
For this purpose it is convenient 
to express the comoving wave number in terms of the present one as
$k_{tod} = \kappa _{tod} T_{\gamma } 
\left( T_I/T_{\gamma }\right) ^{1/b}/HT_{tod}$.
Replacing everything in eq. (\ref{aa}) we obtain
\begin{equation}
\langle B_{\lambda }^2 \rangle \simeq \frac{e^6H^5}{m(0)^2T_{\gamma }^2}
\kappa _{tod}^4 \left( \frac{T_{\gamma }}{T_{tod}} \right) ^4
\left( \frac{T_I}{T_{\gamma }} \right) ^{7/4b}  t_{max} .
\label{er}
\end{equation}

Equation (\ref{er}) gives the intensity of the field at the moment when
the electric current vanishes. After that, the field propagates as
$B_{\lambda}^{phys}(t) = \sqrt{\left< B_{\lambda}^2\right>} 
a\left( t_{max}\right) ^2/a\left( t\right) ^2$ 
(i.e. magnetic flux conservation).
Using the relation $a(t) \propto 1/T$, the present value of the magnetic field
is given by
\begin{equation}
\langle B_{\lambda }^{phys} \rangle \simeq 
\sqrt{\langle B^2_{\lambda } \rangle } 
\left( \frac{T_{tod}}{T_{max}} \right) ^2 \label{es}
\end{equation}
where $T_{max}$ is the temperature of the Universe when the particles decay,
given by
$T_{max} = T_{\gamma } a_{\gamma }\left( \tau _{\gamma }\right)/a_{\gamma }
\left( \tau _{max}\right) = 
T_I^{3/4b}T_{\gamma }^{1-3/4b}/\sqrt{2H t_{max} }$.
Replacing everything in Eq. (\ref{es}) we obtain
\begin{equation}
\langle B_{\lambda }^{phys} \rangle \simeq 
\frac{e^3 H^{7/2} \kappa _{tod}^2}{m(0)T_{\gamma }} 
\left( \frac{T_I}{T_{\gamma }} \right) ^{-5/8b} t_{max}^{3/2}.
\label{eu}
\end{equation}
In the above, we have only considered the effect induced by the scalar
particle and not by the charged particles resulting from its decay.
This might seem surprising
since, due to charge current conservation, the charged particles coming
from the scalar particle decay might also
contribute in a relevant way to the magnetic field generation. However,
the decay of a massive scalar particle, like the stau in the case under study,
will lead mostly to charged fermions (tau leptons, in this case) 
with wavelengths much shorter than the ones of the original scalar
field. These fermions might eventually generate magnetic fields, but,
due to the wavelengths involved, these fields will not be coherent
in the scales of interest for our study.

In order to apply the above formalism to the case of gauge mediated
supersymmetry-breaking models,
we should recall Eq.~(\ref{ev}), which gives the lifetime of
the NLSP, $\tau_{\tilde{\tau}} \equiv t_{max}$
as a function of the supersymmetry-breaking scale and the mass
of the lightest stau.
Replacing Eqs. (\ref{cc})
and (\ref{ev}) into Eq. (\ref{eu}), we obtain
\begin{eqnarray}
\langle B_{\lambda }^{phys}\rangle &\simeq &
\frac{e^3H^{7/2}\kappa ^2_{tod}}{T_{\gamma }^{1/6} 
\left[ H M_{Pl} \right] ^{5/12} \times 
100 \; {\rm GeV}} \nonumber \\
&\times & \left[ \frac1{\tilde k^2} \left( 
\frac{100 \; {\rm GeV}}{m(0)}\right) ^{17/3}
\left( \frac{\sqrt{F}}{100 \; {\rm TeV}}\right) ^4 3\times 10^{11} 
GeV^{-1}\right] ^{3/2}
\label{ew}
\end{eqnarray}
We see that the $b$-dependence has disappeared, i.e. 
the result does not depend on
the details of the reheating period.
Using the equivalence $1\; {\rm GeV}^2 \simeq 10^{20}$ Gauss 
and the numerical estimates
$H \simeq 10^{11}$ GeV, $T_{\gamma } \simeq 10^7$ GeV, 
$\kappa _{tod}\simeq 10^{-38}$ GeV (for a galactic scale of the order
of 1 Mpc), 
$m(0) \simeq 100$ GeV and $\sqrt {F/\tilde k} = 10^6$ GeV, we obtain
\begin{equation}
\langle B_{\lambda }^{phys}\rangle \simeq 10^{-12} \; {\rm Gauss}. 
\label{ex}
\end{equation}
This value of the generated magnetic field is sufficient to seed the
galactic dynamo, being also consistent with the bounds imposed by
the anisotropies in the CMBR and by primordial 
nucleosynthesis~\cite{barrow,grasso}.

In the above we have given results for specific values of $\sqrt{F}$,
$H$ and $T_{\gamma}$, for minimal coupling and for $C_H = 0$,
that is for $\nu = 3/2$. It is interesting to discuss the 
dependence on $\sqrt{F}$, $C_H$, as well as on
departures from minimal coupling. In this case it can be
checked that for small $k_{tod}$, the Bogoliubov coefficients
are given by $\alpha_k \sim \beta_k \sim {\cal O}(1)k^{-\nu}$, 
with $\nu$ given by Eq. (\ref{bbg}).
Considering a stau lifetime
$\tau_{\tilde{\tau}} \equiv t_{max} = 10^n$ GeV$^{-1}$ and 
a stau mass $m_{\tilde{\tau}} \simeq 100$ GeV,
the  value of the physical magnetic field is given by
\begin{eqnarray}
\langle B_{\lambda}^{phys} \rangle & = &
10^{(118/3)(\nu - 3/2) - 106/3 + 3n/2} \;\; {\rm Gauss} \times
\nonumber\\
& &
\left(\frac{10^{7} \; {\rm GeV}}{T_{\gamma}}
\right)^{(7-4\nu)/6} \left(\frac{H}{10^{11} \; {\rm GeV}}\right)^{(25+8\nu)/12}
\left(\frac{\kappa_{tod}}{10^{-38}\; {\rm GeV}}\right)^{(5 - 2 \nu)} .
\label{bphys}
\end{eqnarray}
Acceptable values of the magnetic field,
consistent with the cosmological bounds~\cite{barrow,grasso}
are in the range
\begin{equation}
10^{-9} \; {\rm Gauss} \geq B_{\lambda}^{phys} \geq 10^{-21} \; {\rm Gauss}.
\end{equation}
For the case $\nu = 3/2$, and for the mean values of the parameters 
taken above,
this implies  $158/9\simgt n \simgt 86/9$, or, equivalently,
\begin{equation}
10^{-14} \; {\rm sec} \simlt \tau_{\tilde{\tau}} \simlt 10^{-7} \; {\rm sec} .
\label{magn}
\end{equation}
The variation of the above bounds with the cosmological parameters
can be easily obtained from Eq. (\ref{bphys}).

On the other hand, for arbitrary values of $\nu$, the following bound
is obtained:
\begin{equation}
\frac 32 + \frac{79}{118} - \frac{9}{236}n \geq \nu
\geq \frac 32 + \frac {43}{118} - \frac{9}{236}n .
\end{equation}

In gauge-mediated supersymmetry-breaking
models, for a stau mass $m_{\tilde{\tau}} \simeq 100$ GeV
and values of the gravitino mass
$m_{\tilde{G}} \simlt 1$ keV, the stau lifetime, Eq. (\ref{ev}), 
is such that $11 \simlt n \simlt 16$, or equivalently
\bea
10^{-13} \; {\rm sec} \simlt \tau_{\tilde{\tau}} \simlt 10^{-8} \; {\rm sec}
\;\;\;\;\;\;\; {\rm GMSB\; for\; } m_{\tilde{G}} \simlt 1 \; {\rm keV}.
\label{lftgmsb}
\eea
For a mimimal stau coupling to the curvature and $C_H \ll 1$, that is
for  $\nu = 3/2$, Eq. (\ref{lftgmsb}) is in remarkable
agreement with the values required to generate an acceptable
magnetic field, Eq. (\ref{magn}). 

The bounds on $n$ in gauge mediated supersymmetry
breaking models also imply  bounds on $\nu$
\begin{equation}
1.72 \simgt \nu \simgt 1.25 .
\end{equation}
Comparing this expression with the value of $\nu$ for minimal 
coupling of the scalar field,
$\nu \simeq 3/2\sqrt{1 - 4 C_H/9}$, we obtain that $C_H < 0.68$
in order to generate cosmologically relevant values of the
magnetic field. As follows from Eq.~(\ref{bphys}),
only small modifications of the bound on $C_H$ may
be obtained for different values of the cosmological parameters.

Consider now the departure from minimal coupling. Assuming that
$C_H \ll 1$ we have $\nu \simeq 3/2\sqrt{1 - 16\xi/3}$. The bounds on
the lifetime of the stau are satisfied for 
\begin{equation}
0\leq \xi \simlt 0.06;
\end{equation}
we thus obtained
for a non-negligible interval of coupling values, magnetic
fields of an intensity sufficient for these
to be cosmologically important. 
The upper bounds on $C_H$ and $\xi$ quoted above can only be
obtained for values of $\sqrt{F}$ (or equivalently $n$) such that
the gravitino mass $m_{\tilde{G}}$ is close to 1 keV. 

The results given above were obtained for a 
reheat temperature $T_{\gamma} \simeq 10^7$ GeV. As we emphasized
above, for the range of gravitino masses we are concentrating
on, the most relevant  bound on the reheat temperature comes from
Eq. (\ref{cc}), which assures the consistency of
the whole approach. Larger values of the magnetic fields may be
obtained by lowering the value of the reheat temperature. However, 
the final result for the magnetic field, Eq. (\ref{ew}), depends 
very weakly on the value of the reheat temperature $T_{\gamma}$. 
No relevant   departures from the obtained values would be 
obtained even if  the reheat temperature were as low 
as  $T_{\gamma} \simeq 10^3$ GeV. 

In summary, we have shown that cosmologically relevant magnetic
fields may be generated by a scalar field, minimally coupled
to the curvature, so far its lifetime is bounded by Eq.~(\ref{magn}).
The bounds on the lifetime are in excellent agreement with those
obtained in minimal gauge mediated supersymmetry-breaking
models with the lightest stau as the next-to-lightest supersymmetric
particle, for values of the supersymmetry-breaking scale such that
$m_{\tilde G} \simlt 1$ keV. This conclusion is very weakly dependent
on the assumed values of the cosmological parameters. Moreover,
contrary to many models for magnetic field generation proposed
in the literature, the present one is related to the properties
of the low energy effective theory and these properties can be tested 
in accelerator experiments in the near future~\cite{Wolf}. \\
~\\
{\Large {\bf Acknowledgements}} \\
~\\
We thank A. Riotto, M. Giovannini, M. Shaposhnikov,
G. Chalmers, E. Poppitz and J. Russo for useful discussions. 
C.W. would also like to thank the Aspen Center for Physics where
part of this work has been done.
C.W. was supported in part by the US Department of Energy, Division of High
Energy Physics, under Contract W-31-109-ENG-38. A.K., E.A.C. and 
F.D.M. by Conicet, 
University of Buenos Aires and Fundaci\'on Antorchas.


\newpage
\begin{thebibliography}{}
\bibitem{zeldovich}Ya. B. Zel'dovich, A. Ruzmaikin 
and D. Sokoloff, {\it Magnetic
Fields in Astrophysics}, (Gordon and Breach, New York, 1983).
\bibitem{kronberg}P. P. Kronberg, J. J. Perry and E. L. Zukowski, 
Ap. J. 387 (1992) 528.
\bibitem{wolfe}A. M. Wolfe, K. Lanzetta and A. L. Oren, Ap. J. 388 (1992), 17.
\bibitem{oren} A.L. Oren  and A.M. Wolfe, Ap. J.  445 (1995) 624.
\bibitem{widrow}M. S. Turner and L. M. Widrow, Phys. Rev. D 37 (1988) 2743.
\bibitem{dolgov}A. D. Dolgov, Phys. Rev. D 48 (1993), 2499.
\bibitem{ratra}B. Ratra, Ap. J. 391 (1992), L1 and preprint 
GRP-287/CALT-68-1751 (1991).
\bibitem{veneziano} M. Gasperini, M. Giovannini and G. Veneziano,
Phys. Rev. Lett. 75 (1995) 3796. 
\bibitem{spedalieri}F. D. Spedalieri and F. M. Mazzitelli, 
Phys. Rev. D 52 (1995) 6694.
\bibitem{olinto} B. Cheng and A. V. Olinto, 
Phys. Rev. D 50 (1994) 2421; G. Sigl, 
A. Olinto and K. Jedamzik, Phys. Rev. D 55 (1997) 4582.
\bibitem{otros} K. Enqvist and P. Olesen, Phys. Lett B 329 (1994) 195; 
T. Vachaspati, Phys. Lett B 265 (1991) 258; 
A. Dolgov and J. Silk, Phys. Rev. D 47 (1993) 3144; 
C. Hogan, Phys. Rev. Lett. 51 (1983) 1488;
A. P. Martin and A. C. Davies, Phys. Lett B 360 (1995) 71.
\bibitem{shaposhnikov} M. Giovannini and M. Shaposhnikov,
Phys. Rev. Lett. 80 (1998) 22.
\bibitem{ckm}  E. A. Calzetta, 
A. Kandus and F. D. Mazzitelli, Phys. Rev. D 57
(1998), 7139.
\bibitem{allen} B. Allen, Phys. Rev. D 32, (1985), 3136.
\bibitem{gms}
M.~Dine, W.~Fischler, and M.~Srednicki, \NPB{189}{81}{575};
S.~Dimopoulos and S.~Raby, \NPB{192}{81}{353};
M.~Dine and W.~Fischler, \PLB{110}{82}{227};
M.~Dine and M.~Srednicki, \NPB{202}{82}{238};
M.~Dine and W.~Fischler, \NPB{204}{82}{346};
L.~Alvarez-Gaum\'e, M.~Claudson, and M.~Wise, \NPB{207}{82}{96};
C.R.~Nappi and B.A.~Ovrut, \PLB{113}{82}{175};
S.~Dimopoulos and S.~Raby, \NPB{219}{83}{479}.
\bibitem{at1} M.~Dine and A.E.~Nelson, \PRD{48}{93}{1277};
M.~Dine, A.E.~Nelson and Y.~Shirman,
\PRD{51}{95}{1362};
M.~Dine, A.E.~Nelson, Y.~Nir and Y.~Shirman, \PRD{53}{96}{2658}.
\bibitem{wagner} C.E.M. Wagner, Nucl. Phys. B528 (1998) 3.
\bibitem{rpgm} M. Carena, S. Pokorski and C.E.M. Wagner, 
Phys. Lett. B430 (1998) 281.
\bibitem{decay} See, for instance, S. Dimopoulos, M. Dine, S. Raby and
S. Thomas, Phys. Rev. Lett. 76 (1996) 3494; S. Ambrosanio, G.L. Kane,
G.D. Kribs, S.P. Martin and S. Mrenna, Phys. Rev. D54 (1996) 5395.
%
\bibitem{gherghetta98} T. Gherghetta, G.F. Giudice and A. Riotto,
Phys. Lett. B446 (1999) 28.
\bibitem{GTR} G.F. Giudice, I. Tkachev and A. Riotto, JHEP 9908:009 (1999), hep-ph/9907510.
\bibitem{khlebnikov} L.Kofman, A. Linde and A. A. Starobinsky, \PRD{56}{97} 3258;
I. I. Tkachev, \PLB{376}{96}{35}; S. Khlebnikov and I. I. Tkachev, 
\PRL{77}{96} 219.
\bibitem{kolb} E. W. Kolb and M. S. Turner, 
{\it The Early Universe}, (Addison-Wesley
Publ. Co., New York, 1990).
\bibitem{Toni}  D. J. H. Chung, E. W. Kolb and A. Riotto, \PRD{60}{99}: 063504.
\bibitem{Olive} M.K. Gaillard, H. Murayama and K.A. Olive,
Phys. Lett. B355 (1995) 71.
\bibitem{Tkachev} V. Kuzmin and I. Tkachev, Phys. Rev. D59:123006 (1999).
\bibitem{olinto2} B. Cheng and A. Olinto, \PRD{50}{94}{2421};
 G. Sigl, A. Olinto and K. Jedamzik, \PRD{55}{97}{4582};
J. Ahonen, Phys. Lett B382 (1996) 40.
\bibitem{barrow}J. D. Barrow, P. G. Ferreira and J. Silk, 
Phys. Rev. Lett. 78 (1997) 3610;
K. Subramanian and J. D. Barrow, Phys. Rev. Lett 81 (1998) 3575; 
J. Adams, U. H. Danielsson,
D. Grasso and H. Rubinstein, Phys. Lett. B 388 (1996) 253.
\bibitem{grasso}B. Cheng, D. N. Schramm and J. W. Truran, 
Phys. Rev. D 49 (1994) 5006;
A. V. Olinto, D. N. Schramm and J. W. Truran, Phys. Rev. D 54 (1996) 4714;
D. Grasso and H. Rubinstein, Phys. Lett B 379 (1996) 73;
P. J. Kernan, G. D. Starkman and T. Vachaspati, Phys. Rev. D 54 (1996) 7207.
\bibitem{Wolf} See, for example, 
Aleph Collaboration, Phys. Lett. B433 (1998) 1;
Delphi Collaboration, E. Phys. J. C7 (1999) 595;
G. Wolf, talk presented at SUSY99,
http://fnth37.fnal.gov/program4.html.
\end{thebibliography}
\end{document}